\documentclass{iau} 
\usepackage{graphicx}

\title[Magnetic Fields in AGN Jets] 
{Observational View of Magnetic Fields in Active Galactic Nuclei Jets}

\author[Talvikki Hovatta]   
{Talvikki Hovatta
 }

\affiliation{Tuorla Observatory, Department of Physics and
  Astronomy, University of Turku \\V\"ais\"al\"antie 20, 21500 Kaarina, Finland \\ email: {\tt talvikki.hovatta@utu.fi} \\}

\pubyear{2016}
\volume{324}  
\jname{New Frontiers in Black Hole Astrophysics}
\editors{A. Gomboc \& C. Mundell, eds.}
\begin{document}

\maketitle

\begin{abstract}
According to the currently favored picture, relativistic jets in active galactic nuclei (AGN) are
launched in the vicinity of the black hole by magnetic fields extracting energy from the spinning
black hole or the accretion disk. In the past decades, various models from shocks to magnetic 
reconnection have been proposed as the energy dissipation mechanism in
the jets. This paper presents a short review on how linear polarization
observations can be used to constrain the magnetic field structure in the jets of AGN, and how the
observations can be used to constrain the various emission models.
\end{abstract}

\firstsection 
\section{Introduction}
Active galactic nuclei (AGN) are among the most extreme objects in the
universe. They are the centers of distant galaxies hosting a
supermassive black hole, billions of times the mass of the Sun. About
10\% of AGN produce collimated, relativistic outflows or plasma jets
that shine brightly over the entire electromagnetic spectrum from
radio to gamma-ray energies. The formation and stability of these jets
are not yet fully understood. It is believed that energy is extracted
from the central black hole by large-scale magnetic fields
(\cite{blandford77}) while rotating inner regions of the accretion disk
form an outflowing wind that helps to collimate the jet
(\cite{blandford82}). In this picture, the jet is launched as a
magnetically dominated outflow with strong magnetic fields
accelerating the flow to relativistic velocities. General relativistic
magnetohydrodynamic simulations show that the efficiency of
the jet generation can be very high if the black hole is threaded by a
dynamically important magnetic field (\cite{tchekhovskoy11}).

While the simulations suggest that the jets are likely highly
magnetized near the black hole, their composition and magnetic field
structure in the parsec scales, where they produce most of the emission, is much less
clear. A link between the magnetic fields near the black hole and the
parsec-scale jets was established by \cite[Zamaninasab et al. (2014)]{zamaninasab14} who showed
that there is a correlation between the magnetic flux of the jet and
the accretion disk luminosity, as predicted by the magnetically
arrested disk (MAD) model (\cite{tchekhovskoy11}). This result
illustrates how studying the emission regions in the parsec-scale jets
may help us gain knowledge about the jet formation processes.

One of the main open questions is the nature of the emission region. 
Traditionally, the flaring behavior of AGN jets has been well
explained by shock-in-jet models, especially in the radio bands
(e.g., \cite{hughes85, marscher85}). These models assume the jet to
be kinetically dominated at parsec scales. Shock models struggle to
explain the very fast variability detected at high-energy gamma-ray
bands, and models involving magnetic reconnection have been
invoked to explain the high-energy emission
(e.g., \cite{giannios09}). In this case the jet flow should be
magnetically dominated, although recent particle-in-cell simulations
show that even in reconnection models, the magnetic field and particle energies can be in
equipartition at the emission region (\cite{sironi15}).

\begin{figure}[b]
\begin{center}
 \includegraphics[width=\textwidth]{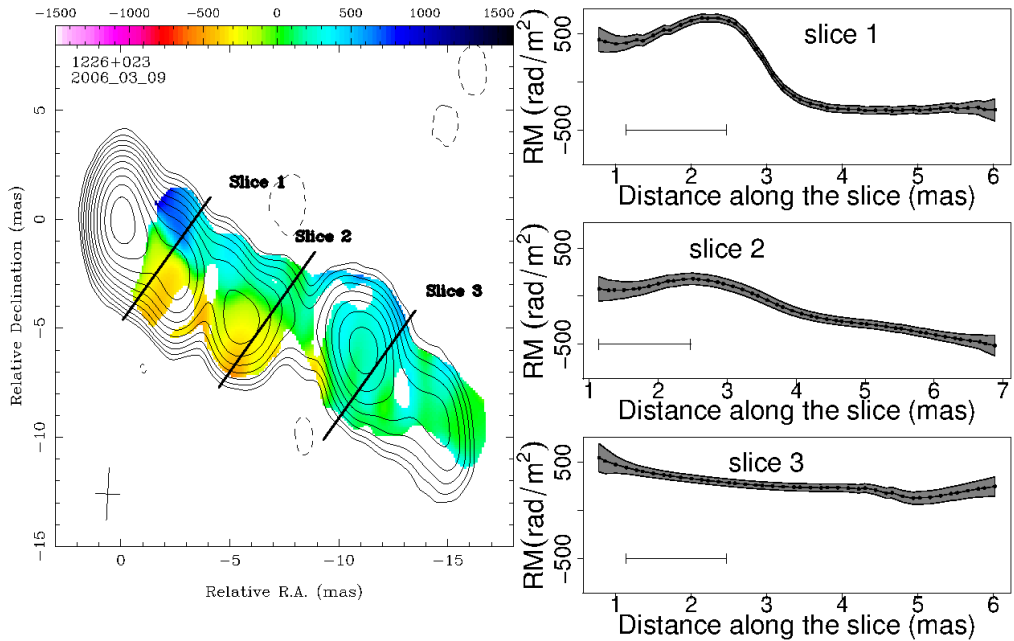} 
 \caption{Faraday rotation map of 3C 273 obtained with the VLBA at 8 -
 15 GHz. A rotation measure gradient is seen extending the full length
of the jet. This can be interpreted as a signature of a helical
magnetic field. Figure adapted from \cite[Hovatta et al. (2012)]{hovatta12}.}
   \label{fig1}
\end{center}
\end{figure}

Another major open question is the structure of the magnetic field at
the emission site. According to the simulations, the magnetic field is
helical close to the black hole but whether it preserves its order in
the parsec scales is debated upon. For example, in the model of
\cite[Marscher et al. (2008)]{marscher08}, the magnetic field is helical at the acceleration
and collimation zone of the jet but gets disrupted and turbulent
beyond a standing shock in the jet. Polarization observations of some
jets, however, indicate that the magnetic field could be helical even
on the parsec scales (e.g., \cite{asada02, gabuzda04,
  hovatta12}). Figure \ref{fig1} shows an example adapted from
\cite[Hovatta et al. (2012)]{hovatta12}.

In this contribution, I will discuss how linear polarization observations in the
radio and optical bands can be used to constrain jet emission
models and the magnetic field structure of the jets. This is not
intended as a comprehensive review of all possible models, but more to
give some examples on how observations are used to study magnetic
fields in AGN jets.

\section{Polarization observations}\label{obs}
The optical and radio emission of AGN jets is synchrotron emission,
which is intrinsically highly polarized. In an optically thin emission
region with a uniform magnetic field,
the polarization degree is up to 70\% (e.g., \cite{pacholczyk70}),
although such high polarization degree values are not typically seen
in AGN jets in the radio (e.g., \cite{aller03, lister05}) or optical
(e.g., \cite{angel80, pavlidou14}) bands. This has been taken as
evidence for disordered magnetic fields. The emission in AGN jets is
often described with the Stokes parameters \textit{I} (for total
intensity), \textit{Q} and \textit{U} (for linear polarization) and
\textit{V} (for circular polarization). I will only discuss linear
polarization here. Using the Stokes parameters, the polarization
degree and the electric vector position angle (EVPA) can be defined as
$m=(\sqrt{Q^2+U^2})/I$ and EVPA$=1/2\tan^{-1}(U/Q)$. In the simplest
  case, in an optically thin jet, the polarization position angle is 
perpendicular to the magnetic field direction and one can use the EVPA observations
to infer the direction of the magnetic field.
However, one should note that in relativistic jets viewed at a small angle to the line of
sight of the observer, the situation is more complicated (\cite{lyutikov05}).

Another way to obtain information about magnetic fields in blazar jets
is via Faraday rotation observations. When synchrotron radiation
passes through magnetized plasma, Faraday rotation of the EVPAs
proportional to the line-of-sight magnetic field and electron density
occurs (e.g. \cite{burn66}). In the simplest case, the effect can
be described by a linear dependence between the observed electric
vector position angle (EVPA; $\chi_\mathrm{obs}$) and wavelength
squared ($\lambda^2$), given by
\begin{equation}\label{eq:RM}
\chi_\mathrm{obs} = \chi_0 + 0.81 \int n_e \mathbf{B} \cdot \mathbf{\mathrm{d}l} = \chi_0 + \mathrm{RM}\lambda^2,
\end{equation}
where $\chi_0$ is the intrinsic EVPA and RM is the rotation measure
(in rad/m$^2$), related to the electron density $n_e$ of the plasma (in cm$^{-3}$) and the magnetic
field component $\mathbf{B}$ (in $\mu$G) along the line of sight (in parsecs). The RM
can thus be estimated by observing the EVPA at different
frequencies. This will give us information on the line-of-sight
component of the magnetic field. For example, if the rotation measure is positive, 
the magnetic field is coming towards the observer, and if negative it
is going away from the observer. 
Therefore it was suggested that a gradient in a Faraday rotation
measure transverse to the jet direction could reveal helical magnetic field structures (\cite{blandford93}). 

\section{Constraining emission models through polarimetry}
Polarization observations of AGN jets in the radio and optical bands
have been conducted since the 1970s (see e.g., \cite{saikia88}
for a review). At the same time, a large number of theoretical models
were developed to explain the total intensity 
and polarization observations of AGN jets (e.g, \cite{jones77,
blandford79, laing80}). In the 1980s it was suggested
that most of the radio variability, especially in the cm-band, are due
to shocks in the relativistic jets where the magnetic field is
predominantly turbulent (\cite{hughes85, marscher85, jones85}). 

The field of polarization modeling has received a new boost since an
exciting connection to high-energy gamma-ray emission was found
(\cite{marscher08}). A rotation in the optical polarization angle
was coincident with an ejection of a new very 
long baseline interferometry (VLBI) component from the core, and a
very-high-energy (VHE; $>$ 100 GeV) detection of the source BL Lac by
the MAGIC telescope. A further connection between the optical
polarization and gamma-ray flares was suggested in two quasars 3C~279
(\cite{abdo10}) and PKS~1510-089 (\cite{marscher10, aleksic14}). In both of
these the optical polarization angle was seen to rotate by more than
180 degrees over a course of 20 to 50 days during which a sharp
gamma-ray flare was observed. Following all these observations, a
number of new models have been published in the last years 
(e.g., \cite{marscher14, zhang14, hughes15, zhang15}). I will go through some of these models
with the emphasis on the observations that can be used to constrain
them.

\subsection{Turbulence in the jets}
Turbulent magnetic field produces stochastic variations to the
polarization angle and degree (\cite{jones85, marscher14}), such
as an EVPA rotation of any length, typically accompanied by a low
polarization degree. The challenge in constraining these types of
models is the stochastic nature of the variations so that it is not
possible to fit the individual light curves
(\cite{marscher14}). Instead, one should observe a large number of
objects and compare the statistical properties of the variations with
the turbulent models.

To study the connection between high-energy emission and optical
polarization in a statistical manner, the RoboPol program was initiated
in 2013 (\cite{pavlidou14}). The RoboPol instrument is mounted on the
1.3-m telescope in Skinakas Observatory in Crete, where the observing
season lasts from April until November. During its first three observing
seasons, RoboPol observed about 100 AGN twice per each week in search
of optical polarization angle rotations. The main goal was to study
the differences in gamma-ray detected and non-detected objects, and to
search for a connection between rotations and flaring behavior.

RoboPol has detected 40 optical polarization angle rotations during
the first three observing seasons, tripling the number of known rotations (\cite{blinov15, blinov16a,
blinov16b}). Comparing the rotations with random walk simulations showed
that while stochastic variability for any individual rotation could not
be ruled out, it was highly unlikely that all of them would be of
random walk origin. \cite[Kiehlmann et al. (2016)]{kiehlmann16} suggested that the smoothness
of the rotation could be used as an additional indicator when
comparing the rotations with random walk models. They also illustrated the
importance of good sampling in polarization observations by revealing
how additional data changed the rotation reported by
\cite[Abdo et al. (2010)]{abdo10} in the quasar 3C~279.  

\begin{figure}[b]
\begin{center}
 \includegraphics[width=3.4in, angle=-90]{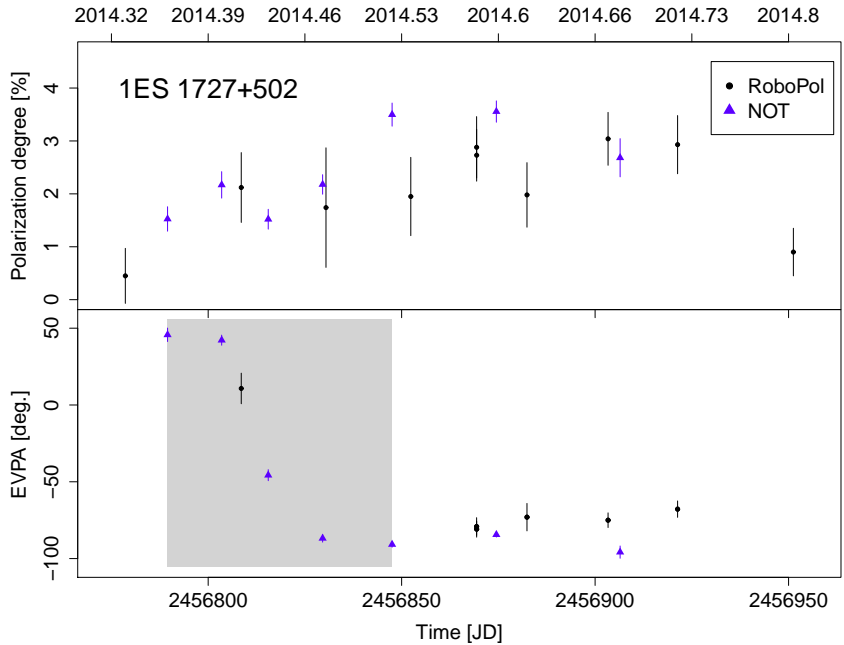} 
 \caption{Polarization degree and EVPA of 1ES~1727+502 in 2014
   observed with RoboPol and the Nordic Optical Telescope (NOT). The
   EVPA shows a rotation of nearly 150 degrees in about 60 days. Figure adapted from \cite[Hovatta et al. (2016)]{hovatta16}.}
   \label{fig2}
\end{center}
\end{figure}

Figure \ref{fig2} shows an example of a EVPA rotation in the source
1ES~1727+502 studied in \cite[Hovatta et al. (2016)]{hovatta16}. Based
on random walk simulations, conducted in the same manner as in
\cite[Kiehlmann et al. (2016)]{kiehlmann16}, a stochastic origin for
the rotation cannot be excluded.

\subsection{Emission in a helical field}
A rotation in the polarization angle could also be due to an emission
feature tracing a magnetic field in the jet as suggested by
\cite[Marscher et al. (2008)]{marscher08} and \cite[Marscher et al. (2010)]{marscher10}. In this model, the magnetic
field in the jet is helical in the acceleration and collimation zone,
which is probed by the optical band, and a rotation is seen when the
emission feature moves along the magnetic field. The rotation may be
accompanied by flaring in other bands when the emission feature
reaches a standing shock in the jet. 

Another alternative is a shock moving down a jet with a helical field,
in which case the EVPA rotation would be due to light travel time effects
when parts of the shock are seen at different times
(\cite{zhang14}). This model was used to successfully fit the EVPA rotation in the
quasar 3C~279 (\cite{zhang15}), although one should bear in mind the
caveat that with the additional data from \cite[Kiehlmann et al. (2016)]{kiehlmann16}, the
rotation was not as long as originally presented by
\cite[Abdo et al. (2010)]{abdo10}. 

Whether the magnetic field in the jet is helical, especially in
regions beyond the acceleration and collimation zone, is still
unclear. As explained in Sect.~\ref{obs}, observations of a Faraday
rotation measure gradient across the jet could be an indication of
such a field. First signatures of a helical magnetic field were
observed in the quasar 3C 273 using multifrequency very long baseline
array (VLBA) observations (\cite{asada02, zavala05}). 
Many more claims of such gradients have been made (e.g., \cite{gabuzda04}) but the issue has been controversial due to the 
limited resolution across the jets in blazars (\cite{taylor10}). 

In \cite[Hovatta et al. (2012)]{hovatta12}, we performed Monte Carlo simulations to quantify
the significance of the rotation measure gradients in the large sample
of parsec-scale jets in the MOJAVE (Monitoring of Jets in Active
galactic nuclei with VLBA Experiments) sample. Our observations
confirmed the gradient in 3C~273 (see Fig.~\ref{fig1}), and we also found significant
gradients in three other quasars. In \cite[Zamaninasab et al. (2013)]{zamaninasab13} we modelled
the gradient of the quasar 3C~454.3 using a magnetic field with both
helical and turbulent components. In the other two sources, the
gradient span only over a small portion of the jet and detailed
modeling was not possible. 

Nowadays performing simulations to quantify the significance of the
gradients is a common practice (e.g., \cite{murphy13}) and the number
of significant gradients has steadily increased (e.g.,
\cite{gabuzda15}). However, rotation measure gradients can also arise from
changes in the density of the Faraday rotating material
(e.g. \cite{gomez11}) so that
detailed modeling of the gradients should be done in order to confirm
that they are indeed due to helical fields. It is also unclear whether
the helical field is within the jet or in a sheath layer around
the jet because Faraday rotation is a propagation effect originating
in the plasma outside the emission region.

\subsection{Shock in a jet}
As stated earlier, shock-in-jet models have been successful in
reproducing the observed flaring especially in the radio
bands. Typically, radiative transfer simulations are generated to
estimate the parameters of the shocks (e.g. \cite{hughes11, hughes15})
and these are then compared with observations at multiple bands
(\cite{aller14}). For example, as shown by \cite{hughes15}, 
the polarization degree and range of EVPA values can 
be used to constrain the magnetic field geometry and jet orientation, 
while total intensity behavior is indistinguishable in different models. 

Interestingly, although the magnetic field is assumed to be predominantly
turbulent, in some cases an ordered field component (possibly helical)
could also be present (\cite{aller16}). This supports the 
findings of e.g., \cite[Zamaninasab et al. (2013), Blinov et al. (2015)]{zamaninasab13, blinov15}, and
\cite[Kiehlmann et al. (2016)]{kiehlmann16}, who also suggest that there are both turbulent
and ordered magnetic field components / deterministic variations in the jets. 

\subsection{Magnetic reconnection}
Magnetic reconnection (see \cite{kagan15} for a review) is a new hot topic in the field. 
It was originally invoked to explain the fast high-energy variability
of blazars through the jet-in-jet model (\cite{giannios09}). In the
recent years, particle-in-cell simulations have evolved significantly,
and it is now possible to reliably simulate the complicated structure
of the reconnection layer (e.g., \cite{sironi15}). While the
reconnection models can already be compared to total intensity
variability of the sources (\cite{petropoulou16}), there are no
explicit models for the polarized emission.

\cite[Zhang et al. (2015)]{zhang15} stated that the EVPA swing in the 2009 flare of 3C~279
was reproduced by a model that ``favors magnetic energy dissipation
process during the flare'' because the magnetic field strength was
seen to decrease when the total intensity increased. However, this
model did not yet include detailed comparisons to a specific
reconnection model. Considering the growing interest in the magnetic
dissipation models, it is likely that in the next few years more
detailed models with observational predictions will become available.

\subsection{Statistical trends}
From the previous sections, it is clear that polarization observations
can be used to constrain various types of emission models and the
magnetic field structure in the jets. In order to generalize the
findings in individual sources into the AGN jet population, a statistical approach must be
used. This is what the RoboPol program aims for by observing a sample
of about 100 sources with high cadence. RoboPol has found, for
example, that any class of blazars from low to high synchrotron
peaking objects can show rotations (see
also \cite{jermak16, hovatta16}). However, there seems to be a
specific class of ``rotators'' that do so more often, and rotations
are more common in the low spectral peaking objects (\cite{blinov16b}).

The low synchrotron peaking objects also have higher optical polarization degree
than the high-peaking objects (\cite{angelakis16}), a trend that has
also been seen in the smaller sample studied in \cite[Jermak et al. (2016)]{jermak16} and
earlier in the radio band (\cite{lister11}). These general trends can
be explained with a simple, qualitative model where a shock moves down
a jet, which has both helical and turbulent magnetic field components (see
\cite{angelakis16} for details). Any model put forward to explain the
polarization behavior in individual flares, should also account for
these general trends.

\section{Future directions}
One challenge in connecting the observations of magnetic fields to the
theory of jet formation is that often the observations, especially in
the radio bands,  probe regions $10^3-10^5$ gravitational radii away from the black hole. 
Optical observations do not suffer from this restriction, but for
example, Faraday rotation with $\lambda^2$ wavelength dependence, is
not typically seen at optical wavelengths. A solution can be provided by going
to millimeter-band observations, as demonstrated by
\cite[Mart\'i-Vidal et al. (2015)]{marti-vidal15}. They used ALMA polarization observations to
probe the Faraday rotation at the jet base of the lensed quasar PKS~1830-211, and found extremely high Faraday rotation of
$10^8$rad/m$^2$, which is two orders of magnitude higher than previous
observations in other sources (e.g., \cite{plambeck14}). They inferred
this as a signature of a very high
magnetic field at the base of the jet, in support of magnetically
launched jet models. Whether similar high RM values are a common
property of quasars remains to be seen with future ALMA
observations. Especially interesting will be future observations using
ALMA as part of the global VLBI array, which may be able to spatially resolve
the regions of high magnetic fields.

A major leap forward will come with the future X-ray polarization
missions because the amount of polarization for different high-energy emission
mechanisms in AGN jets, for example between leptonic and hadronic
models, is very different (\cite{zhang13}), and X-ray polarization
observations can be used to distinguish between the models.

\section{Summary}
Magnetic fields in AGN jets can be probed through polarization
observations, which are most easily done at radio and optical
bands. Modeling the polarization degree and position angle behavior
can be used to test different blazar emission models, while Faraday
rotation observations at radio, and now also in the millimeter bands,
can be used to probe the line-of-sight magnetic field component. A
statistical approach is favored when connecting the results to
different AGN jet populations.


\begin{thebibliography}{}
\bibitem[Abdo et al. 2010]{abdo10}
{Abdo, A. A., Ackermann, M., Ajello, M. et al.} 2010, 
\textit{Nature}, 463, 18

\bibitem[Aleksi\'c et al. 2014]{aleksic14}
{Aleksi\'c, J., Ansoldi, S., Antonelli, L. A. et al.} 2014
\textit{A\&A}, 569, 46

\bibitem[Aller et al. 2003]{aller03}
{Aller, M. F., Aller, H. D. \& Hughes, P. A.} 2003, 
\textit{ApJ}, 586, 33

\bibitem[Aller et al. 2014]{aller14}
{Aller, M. F., Hughes, P. A., Aller, H. D. et al.} 2014, 
\textit{ApJ}, 791, 51

\bibitem[Aller et al. 2016]{aller16}
  {Aller, M. F., Hughes, P. A., Aller, H. D., et al.} 2016,
\textit{Galaxies}, 4, 35

\bibitem[Angel \& Stockman 1980]{angel80}
{Angel, J. R. P. \& Stockman, H. S.} 1980, 
\textit{ARA\&A}, 18, 321

\bibitem[Angelakis et al. 2016]{angelakis16}
{Angelakis, E., Hovatta, T., Blinov, D. et al.} 2016
\textit{MNRAS}, 463, 3365

\bibitem[Asada et al. 2002]{asada02}
{Asada, K. Inoue, M., Uchida, Y., et al.} 2002, 
\textit{PASJ}, 54, L39

\bibitem[Blandford 1993]{blandford93}
{Blandford, R.} 1993, in:  M. Burgarella, M. Livio, \&
C. P. O’Dea (eds.)
\textit{Astrophysical Jets}, (Astrophysics and Space Science Library, Vol. 103; Cambridge:
Cambridge Univ. Press), 15

\bibitem[Blandford \& K\"onigl 1979]{blandford79}
{Blandford, R. D. \& K\"onigl, A.} 1979, 
\textit{ApJ}, 232, 34

\bibitem[Blandford \& Payne 1982]{blandford82}
{Blandford, R. D. \& Payne, D. G.} 1982, 
\textit{MNRAS}, 199, 883

\bibitem[Blandford \& Znajek 1977]{blandford77}
{Blandford, R. D. \& Znajek, R. L.} 1977, 
\textit{MNRAS}, 179, 433

\bibitem[Blinov et al. 2015]{blinov15}
{Blinov, D., Pavlidou, V., Papadakis, I. et al.} 2015, 
\textit{MNRAS}, 453, 1669

\bibitem[Blinov et al. 2016a]{blinov16a}
{Blinov, D., Pavlidou, V., Papadakis, I. E. et al.} 2016a, 
\textit{MNRAS}, 457, 2252

\bibitem[Blinov et al. 2016b]{blinov16b}
{Blinov, D., Pavlidou, V., Papadakis, I. E. et al.} 2016b, 
\textit{MNRAS}, 462, 1775

\bibitem[Burn 1966]{burn66}
{Burn, B. J.} 1966, 
\textit{MNRAS}, 133, 67

\bibitem[Gabuzda et al. 2015]{gabuzda15}
{Gabuzda D. C., Knuettel, S. \& Reardon, B.} 2015
\textit{MNRAS}, 450, 2441

\bibitem[Gabuzda et al. 2004]{gabuzda04}
{Gabuzda, D. C., Murray, E., \& Cronin, P.} 2004, 
\textit{MNRAS}, 351, L89

\bibitem[Giannios et al. 2009]{giannios09}
{Giannios, D., Uzdensky, D. A. \& Begelman, M. C.} 2009, 
\textit{MNRAS}, 359, L29

\bibitem[G\'omez et al. 2011]{gomez11}
{G\'omez, J.-L., Roca-Sogorb, M., Agudo, I., et al.} 2011, 
\textit{ApJ}, 733, 11

\bibitem[Hovatta et al. 2012]{hovatta12}
{Hovatta, T., Lister, M. L., Aller, M. F. et al.} 2012, 
\textit{AJ}, 144, 105

\bibitem[Hovatta et al. 2016]{hovatta16}
{Hovatta, T., Lindfors, E., Blinov, D. et al.} 2016
\textit{A\&A}, 596, 78

\bibitem[Hughes et al. 1985]{hughes85}
{Hughes, P. A., Aller, H. D. \& Aller, M. F.} 1985, 
\textit{ApJ}, 298, 301

\bibitem[Hughes et al. 2011]{hughes11}
{Hughes, P. A., Aller, M. F., \& Aller, H. D.} 2011, 
\textit{ApJ}, 735, 81

\bibitem[Hughes et al. 2015]{hughes15}
{Hughes, P. A., Aller, M. F., Aller, H. D.} 2015, 
\textit{ApJ}, 799, 207

\bibitem[Jermak et al. 2016]{jermak16}
{Jermak, H., Steele, I., Lindfors, E. et al.} 2016,
\textit{MNRAS}, 462, 4267

\bibitem[Jones \& O'Dell 1977]{jones77}
{Jones, T. W. \& O’Dell, S.} 1977, 
\textit{ApJ}, 215, 236

\bibitem[Jones et al. 1985]{jones85}
{Jones, T. W., Rudnick, L., Aller, H. D. et al.} 1985, 
\textit{ApJ}, 290, 627

\bibitem[Kagan et al. 2015]{kagan15}
{Kagan, D., Sironi, L. Cerutti, B. \& Giannios, D.} 2015, 
\textit{Space Science Reviews}, 191, 545

\bibitem[Kiehlmann et al. 2016]{kiehlmann16}
{Kiehlmann, S., Savolainen, T., Jorstad, S. G., et al.} 2016, 
\textit{A\&A}, 590, A10

\bibitem[Laing 1980]{laing80}
{Laing, R. A.} 1980, 
\textit{MNRAS}, 193, 439

\bibitem[Lister \& Homan 2005]{lister05}
{Lister, M. L., \& Homan, D. C.} 2005, 
\textit{AJ}, 130, 1389

\bibitem[Lister et al. 2011]{lister11}
{Lister, M. L., Aller, M. F., Aller, H. D. et al.} 2011,
\textit{ApJ}, 742, 27

\bibitem[Lyutikov et al. 2005]{lyutikov05}
{Lyutikov, M., Pariev, V. I., \& Gabuzda, D. C.} 2005, 
\textit{MNRAS}, 360, 869

\bibitem[Marscher 2014]{marscher14}
{Marscher, A. P.} 2014, 
\textit{ApJ}, 780, 87

\bibitem[Marscher \& Gear 1985]{marscher85}
{Marscher, A. P. \& Gear, W. K.} 1985, 
\textit{ApJ}, 298, 114

\bibitem[Marscher et al. 2008]{marscher08}
{Marscher, A. P., Jorstad, S. G., D'Arcangelo, F. D. et al.} 2008, 
\textit{Nature}, 452, 966

\bibitem[Marscher et al. 2010]{marscher10}
{Marscher, A. P., Jorstad, S. G., Larionov, V. M. et al.} 2010, 
\textit{ApJ}, 710, L126

\bibitem[Mart\'i-Vidal et al. 2015]{marti-vidal15}
{Mart\'i-Vidal, I., Muller, S., Vlemmings, W., Horellou, C., Aalto,
  S.} 2015, 
\textit{Science}, 348, 311

\bibitem[Murphy \& Gabuzda 2013]{murphy13}
{Murphy  E.  \&  Gabuzda  D. C.}  2013,  
in \textit{The  Inner-most  Regions  of  Relativistic  Jets  and  Their  Magnetic
Fields}, EPJ Web of Conferences, Volume 61, id.07005

\bibitem[Pacholczyk 1970]{pacholczyk70}
{Pacholczyk, A. G.} 1970, 
\textit{Radio Astophysics, Nonthermal Processes in Galactic
and Extragalactic Sources} (San Francisco, CA: Freeman)

\bibitem[Pavlidou et al. 2014]{pavlidou14}
{Pavlidou, V., Angelakis, E., Myserlis, I. et al.} 2014, 
\textit{MNRAS}, 442, 1693

\bibitem[Petropoulou et al. 2016]{petropoulou16}
{Petropoulou, M., Giannios, D., Sironi, L.} 2016, 
\textit{MNRAS}, 462, 3325

\bibitem[Plambeck et al. 2014]{plambeck14}
{Plambeck, R. L., Bower, G. C., Rao, R. et al.} 2014, 
\textit{ApJ}, 797, 66

\bibitem[Saikia \& Salter 1988]{saikia88}
{Saikia, D. \& Salter, C.} 1988, 
\textit{Ann. Rev. in Astron. \& Astrophys.} , 26, 93

\bibitem[Sironi et al. 2015]{sironi15}
{Sironi, L., Petropoulou, M., \& Giannios, D.} 2015, 
\textit{MNRAS}, 450, 183

\bibitem[Taylor \& Zavala 2010]{taylor10}
{Taylor, G. B., \& Zavala, R. T.} 2010, 
\textit{ApJ}, 722, L183

\bibitem[Tchekhovskoy et al. 2011]{tchekhovskoy11}
{Tchekhovskoy, A., Narayan, R., McKinney, J. C.} 2011, 
\textit{MNRAS}, 418, L79

\bibitem[Zamaninasab et al. 2014]{zamaninasab14}
{Zamaninasab, M., Clausen-Brown, E., Savolainen, T. \& Tchekhovskoy,
  A.}  2014, 
\textit{Nature}, 510, 126

\bibitem[Zamaninasab et al. 2013]{zamaninasab13}
{Zamaninasab, M., Savolainen, T., Clausen-Brown, E. et al.} 2013, 
\textit{MNRAS}, 436, 3341

\bibitem[Zavala \& Taylor 2005]{zavala05}
{Zavala, R. T., \& Taylor, G. B.} 2005, 
\textit{ApJ}, 626, L73

\bibitem[Zhang \& B\"ottcher 2013]{zhang13}
{Zhang, H. \& Böttcher, M.} 2013, 
\textit{ApJ}, 774, 18

\bibitem[Zhang et al. 2014]{zhang14}
{Zhang, H. Chen, X. \& B\"ottcher, M.} 2014, 
\textit{ApJ}, 789, 66

\bibitem[Zhang et al. 2015]{zhang15}
{Zhang, H. Chen, X. \& B\"ottcher, M. et al.} 2015, 
\textit{ApJ}, 804, 58

\end{thebibliography}
\end{document}